# Spectral representation – analyzing single-unit activity in extracellularly recorded neuronal data without spike sorting

Artur Luczak *

Center for Molecular and Behavioral Neuroscience,

Rutgers, The State University of New Jersey, Newark, NJ.

Dept of Computer Science, Yale University, New Haven, CT.

Nandakumar S. Narayanan

John B Pierce Laboratory, New Haven, CT.

Interdepartmental Neuroscience Program, Yale University, New Haven, CT.

April 2004

\_\_\_\_\_\_\_\_\_\_.

* To whom correspondence should be addressed: E-mail: Luczak@cs.yale.edu

**Abstract**

One step in the conventional analysis of extracellularly recorded neuronal data is spike sorting, which separates electrical signal into action potentials from different neurons. Because spike sorting involves human judgment, it can be subjective and time intensive, particularly for large sets of neurons. Here we propose a simple, automated way to construct alternative representations of neuronal activity, called spectral representation (SR).  In this approach, neuronal spikes are mapped to a discrete space of spike-waveform features and time. Spectral representation enables us to find single-unit stimulus related changes in neuronal activity without spike sorting. We tested the ability of this method to predict stimuli using both simulated data and experimental data from an auditory mapping study in anesthetized marmoset monkeys. We find that our approach produces more accurate classification of stimuli than spike sorted data for both simulated and experimental conditions. Furthermore, this method lends itself to automated analysis of extracellularly recorded neuronal ensembles. Additionally, we suggest ways in which these representations can be readily extended to assist in spike sorting and the evaluation of single-neuron peri-stimulus time histograms.

**1. Introduction**

Spike sorting, an important step in the analysis of extracellularly recorded neuronal data, relies on proper assignment of spikes to neurons in order to draw inferences from neuronal recordings.  Many methods have been proposed for spike sorting (for review see [15]; selected recently proposed methods: [6, 7, 10, 11, 14, 18, 20, 21, 26]). Typically, the researcher labels the waveforms as belonging to one or another neuron based on the spike waveforms. The selection of criteria for spike sorting is heavily operator dependent and consequently subjective and time consuming. Spike sorting can be especially difficult when the signal to noise ratio is low or when there are non-stationarities within the neuronal signal, such as variations in background neuronal activity. Furthermore, the error rate of spike sorting usually exceeds 20 % [9, 25]. Additionally, as ensemble recordings gain in popularity, spike sorting becomes a limiting step in the analysis of neuronal data, particularly when several hundred neurons are recorded in a single experiment [17].

Here, we propose a new method for the analysis of extracellularly recorded neuronal ensemble data, which does not require traditional spike sorting. In this method, we construct discrete spectral representation (SR) of neuronal signals in time-feature space and apply pattern recognition methods to determine task or stimulus-related changes in SR. This representation is analogous to spectrograms of sound, where one-dimensional sound waveform is mapped to time – frequency (sound waveform features) space. This method is intended to facilitate prediction of stimuli from neuronal responses by identifying stimulus-related changes in the time course of features of neuronal activity without assigning these features to distinct neurons (i.e. stimulus prediction without spike sorting). It is similar to detecting differences between two pieces of music by comparing sound spectrograms without assigning sound frequencies to distinct music notes (analogy to spike sorting). Importantly, in this method information about neurons' identities is not lost, as in the case of the analysis of multi-unit activity. When it is of interest to use spike-sorted signals, the SR can also be applied to assist in analyzing the remaining multi-unit activity signal.

We apply the SR method to simulated data and demonstrate the advantage of this method over standard spike sorting analysis when neurons with very similar waveforms could not be correctly clustered. We then apply the SR method to extracellularly recorded neuronal data from auditory cortex in anesthetized marmoset monkeys. We demonstrate that the SR approach produces accurate classification of acoustic stimuli presented to the monkey on single trial basis. Finally, we demonstrate applications of the SR approach in creating single-unit peri-stimulus time histograms, and for 'spectral spike sorting', or the identification of activity related to individual neurons.

## 2. Methods
### *2.1. Constructing Spectral Representations (SR)*

In our method, as in many spike sporting approaches, spike waveforms are represented by their features (Fig. 1). However, unlike spike sorting methods, the SR method does not cluster spike waveforms, which is the most difficult step in traditional spike sorting. In order to derive spike features we used principal component analysis (PCA). As a result, spike waveforms were represented by a few numbers describing different features of spikes. Using these features, every spike was represented as a point in $N+1$ dimensional space: $p = (pc_1, pc_2, ..., pc_N, t)$, where $pc_1,...,pc_N$ are the PC scores of a spike waveform, $N$ is a number of PCs, and $t$ is peri-stimulus time of spike. This is equivalent to representing a spike by $N$ points in $N$ two-dimensional spaces: $p_1 = (pc_1, t)$, $p_2 = (pc_2, t),..., p_N = (pc_N, t)$. The span of values of $pc_1,..., pc_N$ can be binned in $M$ intervals and mapped (by scaling and translation) to $M$ integers on interval 1-$M$. Time was binned into $C$ intervals. After binning, two-dimensional spaces were replaced by two-dimensional arrays with $M$ rows and $C$ columns. All arrays were concatenated to form one two-dimensional SR array with $N$x$M$ rows and $C$ columns. In the SR array every spike was represented by $N$ 'ones' corresponding to points: $p_1, p_2, ..., p_N$. In order to improve estimated distribution of spikes, the SR array was convolved with a smoothing kernel. Figure 1 provides a graphical illustration of the mapping of neuronal signal (panel A) in an SR array (panel B). 'Ones' in figure 1B represents point $p$ and fractions around 'ones' reflect smoothing. A sample calculation of the mapping is provided in the appendix.

The number of points (i.e. PCs) used to describe spike waveform were determined from magnitudes of eigenvalues, e.g. from a scree-plot (example of a scree-plot is in Fig. 4D). Note that the SR array can be readily expanded by adding rows with values describing a variety of time-varying features of stimulus or animals behavior. For instance, an SR array can contain information about neuronal signal and an additional stimulus related variable, such as sound direction in an auditory stimuli discrimination task (Luczak et al. [16] describes a similar approach for receptive field mapping).

*2.2. Stimulus discrimination based on SR of neuronal signals*

For single-trial discrimination of stimuli we constructed SR arrays from neuronal data. The partial least squares (PLS) method was used to determine stimulus-relevant features in SR. PLS is a similar method to PCA. The difference between the two is that PCA finds directions of greatest variability in data whereas PLS finds directions of greatest variability correlated with stimulus. Alternative methods, e.g. wavelet-based discriminate pursuit also can be used [2, 13]. Note that the feature selection step also serves as a denoising procedure. After feature selection, linear discriminate analysis (LDA) was used to determine the stimulus type. Leave-one-out cross validation was used in feature selection and classification. We used these analysis techniques on a sample simulation to illustrate stimulus discrimination (see the Results section).

*2.3. Estimating firing patterns of single-units without spike sorting*

The SR array encodes information about spike times as well as spike features. Therefore, it is possible to evaluate the activity of a neuron with specific spike features directly from the SR array. For example, PCA analysis of the SR array results in principal components, which describe changes in the temporal structure of SR. The first PC of the SR array describes mean neuronal activity (equivalent to multi-unit PSTH). The second PC describes the largest variability in SR after subtracting the mean activity from SR array. Thus, the second PC can be interpreted as the activity of the neuron with the highest firing rate subtracted from the activity of other neurons. The subsequent PCs can be interpreted analogously as neural activity with subsequent firing rate subtracted from the activity of neurons not accounted by previous PCs. In other words, PCA evaluates differences between neurons activity (Fig 4B). In order to approximate single units PSTHs, independent component analysis (ICA) or factor analysis (FA) can be used instead of PCA. Unfortunately, when using ICA or FA it is difficult to achieve a stable solution for high dimensional data.

*2.4. Spectral spike sorting.*

In previous section we were interested in changes in neuronal activity along the time axis. For assigning features to individual neurons, the problem is 'transposed'. We are interested in finding distinct distributions along the spike features axis. One approach is to apply PCA to the transposed SR array (see example in Fig. 4C). The first PC describes the mean distribution of features. For easily separable spike waveforms, the first PC would have distinct maxima. The subsequent PCs describe temporal changes in the distribution of spike features. Thus, it is possible to distinguish among neurons with similar spike waveforms, but with different tuning properties. The parts of the SR array that correspond to the same sign of the PC are positively correlated. Therefore, the point at which the PC is changing its sign can be regarded as the demarcation line between waveform clusters [23].

Peri-stimulus time can provide substantial information for spike sorting when neurons have similar spike waveforms but different tuning properties or respond with different time delay after a stimulus (a similar idea was used to identify motor unit action potentials [22]). Because the SR array provides information about peri-stimulus time and waveforms, the SR array is more advantageous for assigning features to individual neurons than classical approaches.

In order to validate spectral assignment of features to individual neurons, this method was tested on artificial and real data as described in the Results section.

*2.5. Simulated Data*

In order to demonstrate and validate the use of SR, artificial data were created. We simulated two neurons (denoted by A and B) with similar extracellular spike waveforms but with distinct firing patterns. The spikes waveforms features had normal distribution with unit variance in PCA features space (Fig. 2). Every spike waveform was described by two principal components. Thus, each spike was represented by two points in the SR array (Fig. 3D, E). The maximal density of spikes was formed in the middle point between the centers of distributions (Fig. 2B, C solid line).

We simulated neuronal responses to two stimuli. In response to Stimulus 1 neuron A was active between 0-100 ms, and neuron B between 50-150 ms (Fig. 3A). Stimulus 2 had the opposite response: neuron A was active between 50-150 ms and neuron B was active between 0-100 ms (Fig. 3A). For both stimuli the multiunit activity was exactly the same (Fig. 3C). For each stimulus 40 trials were simulated with twenty spikes per trial (ten spikes per each neuron).

*2.6. Electrophysiology*

Neuronal data were kindly provided by Troy A. Hackett and Yoshinao Kajikawa from Vanderbilt University. All procedures were approved by the IACUC at Vanderbilt University. Neural recordings were obtained in two anesthetized (ketamine hydrochloride, 10 mg/kg and xylazine, 2mg/kg; I.M.) marmoset monkeys (*Callithrix jacchus jacchus*) using standard neurophysiological methods. A multi-channel recording system (4-channel Bioamp and Brainware software) from Tucker-Davis Technologies (Gainesville, FL) was used. Electrodes were made of Parylene C or Polyimide-coated tungsten, were 3 μm at the tip, and had impedances of 1MΩ at 1kHz. Neuronal spike trains were recorded simultaneously from four electrodes located in the primary and caudal-medial auditory cortex [8]. Data were collected from seventeen separate penetrations. The recordings consisted of clearly resolved single units and multiple unit clusters containing spikes from several neurons that could not be resolved using standard methods for spike sorting (i.e., thresholding and PCA). For the analysis, signals from twenty electrodes locations with twenty-three resolved units were chosen.

*2.7. Acoustic Stimuli*

Experiments were conducted by Drs Troy A. Hackett and Yoshinao Kajikawa in a sound-isolating chamber (Industrial Acoustics Corp., NY) located within the auditory research laboratory at Vanderbilt. Auditory stimuli were presented using Tucker-Davis Technologies (Gainesville, FL) System II hardware and software. Stimuli were calibrated using a ¼" microphone (ACO Pacific, CA) and Tucker-Davis calibration software (SigCal). In this study, we analyzed responses to two frequency sweeps: 2-10 kHz and 6-20 kHz with duration 30 ms and 50 ms respectively. The stimuli were 3.5 s long and consisted of 4 repetitions of the same sweep with 1 s interval between sweeps. Each of the stimuli was presented ten times. The RMS amplitude of the sweeps was adjusted to 60 dB SPL.

## 3. Results
### 3.1. Simulations
#### 3.1.1. Stimulus discrimination

Stimulus discrimination with the SR method using simulated data is illustrated in Figure 3. For Stimulus 1 (Fig. 3A, D) and Stimulus 2 (Fig. 3B, E) the neurons had the opposite responses (Fig. 3A, D). For both stimuli multiunit activity was exactly the same (Fig. 3C). The difference between SRs for both stimuli (Fig. 3F) indicates parts of the SR array, which can be used for stimulus discrimination. Those stimulus-related changes in SR were identified automatically with PLS and LDA was used for single trial predictions. Accuracy of stimulus classification was 92% for SR. Note, that in this example, reliable spike sorting is difficult (see section 3.1.3) and standard neuronal data analysis approaches perform poorly. For instance, using k-mean clustering for spike sorting, resulting in assignment of the center of the first cluster in the middle point between the actual centers of simulated distributions. Classification accuracy for multiunit activity was at the chance level (50%), because multiunit activity does not change with stimulus type.

#### 3.1.2. Estimating firing patterns.

Using the SR approach, we were able to approximate single unit PSTHs directly by decomposing the SR array with PCA. Note that no explicit spike sorting is required to approximate the PSTH from the SR array. The SR array had $C = 100$ columns representing $C$ time intervals, and $N$x$M$ rows, where $N = 2$ is the number of PCs used to describe spike waveforms, and $M = 100$ is the number of PC intervals (bins). Thus, the PCA of an SR array can be interpreted as an analysis of the array containing $N$x$M$ signals, each $C$-dimensional.

The first PC (*pc1*) of the SR array describes mean neuronal activity (multi-unit PSTH; Fig. 4B, solid line). The second PC (*pc2*) describes the difference between the activity of neuron A and neuron B (PSTH of neuron A minus PSTH of neuron B; Fig. 4 B, dashed line). The positive part of *pc2* corresponds to period 0-50 ms when the first unit is active alone. The middle part of *pc2* (50-100 ms) has values around zero, corresponding to the equal activity of both units. The third part of *pc2* (100-150 ms) is negative, reflecting the activity of the second neuron alone.

Note that with proper scaling $pc1 + pc2$ = PSTH of neuron A, and $pc1 - pc2$ = PSTH of neuron B.

#### 3.1.3. Spectral spike sorting

In simulations where distributions of spike-features overlapped by 45%, reliable spike sorting is difficult. The maximal density of spikes is formed in the middle point between the centers of distributions (Fig. 2B,C solid line). For that reason, k-mean clustering, commonly used for spike sorting, fails by definition. The gaussian mixture model also incorrectly located the center of the first cluster in the middle point between the actual centers of distributions (in the same place as k-mean clustering).

In our approach, in order to assign features to individual neurons, we found clusters of spike features by applying PCA to the transposed SR array (Fig. 4C). The first PC (*pc1*) of the transposed SR array (Fig. 4C, solid line) describes mean distribution of spikes features in SR (distributions of *pc1* values is equivalent to histograms in Fig. 2B and C). The second PC (*pc2*) reflects differences in distributions between units (Fig. 4C, dashed line).

As previously described, the SR array extends the standard representation of spikes waveforms in PC-coordinate to the time domain. Hence, differences between units described by *pc2* can be mapped back to standard PC-coordinates (an example of such mapping is illustrated on Fig. 7A, B, C). In figure 4, the point at which *pc2* changes sign (PC1: 0.51 and PC2: 0.48) accurately estimates the borders between the activity of distinct neurons. The optimal boundary between simulated neuronal features is at (PC1: 0.5, PC2: 0.5). These results demonstrate that by PCA decomposition of the SR array we could find borders between units in PC-coordinates. Moreover, the position of extremes of *pc2* (minima at PC1: -0.25, PC2: -0.34 and maxima at PC1: 1.33, PC2: 1.46) approximates the centers of clusters. The actual centers of the simulated distributions were located at (0,0) and (1,1) respectively (SD = 1 for both distributions). In this simulation, because 45% of distributions overlap, points of the highest variability between units were not exactly in the centers of the clusters. Therefore, the location of *pc2* extremes was shifted toward the side of the distribution comparing to location of the actual clusters centers.

The number of recorded neurons can be estimated from the number of principal components necessary to describe the variability present in the SR array, because every unit contributes a unique variability which can be described by a single PC. Figure 4D presents the percentage of variability captured by consecutive PCs. The first two PCs describe 91% of variability. Eigenvalues of subsequent PCs are close to zero, thus indicating that only two units are present in the SR, and that is the correct number of simulated neurons.

The Matlab code to reproduce the simulations is available upon request.

### *3.2. Spectral analysis of auditory neurons from marmoset monkey.*

In this study, we analyzed neuronal responses to two frequency sweeps starting at 2 kHz and 6 kHz respectively recorded from the auditory cortex of marmoset monkeys. The spike waveforms were analyzed with PCA. Based upon the analysis of eigenvalues, only the first two principal components were used to represent spikes. We used 1 ms time bins. Additionally, 5 ms and 10 ms time bins were used to confirm presented results (data not shown). Figure 6A shows the SR of neuronal responses to forty 2 kHz sweep (10 presentations of 4 sweeps series).

### *3.2.1. Stimulus discrimination*

Using PLS for feature selection in SR, and LDA for discrimination, the mean correct classification rate of single trial data was 80.1 %. For comparison, the classification rate of spike-sorted signals was 73.6 % (Fig. 5A). The SR approach achieved a significantly higher prediction rate for presented acoustic stimuli (paired t-test, t = 0.012, P < 0.05).

An example of the SR of neuronal responses to frequency sweeps is illustrated in figure 5B. The difference between the SR arrays representing both stimuli is shown in figure 5C (only part of SR array corresponding to PC-1 and 0-40 ms after stimulus onset is shown). The positive peak followed by the negative part corresponds to the longer response time to 6 kHz sweeps. The largest modulations in the lower part of the SR indicate that the neuron with spikes described by low values of PC-1 is the most discriminative for presented sweeps. This illustrates extraction of information about tuning properties of a single neuron from the SR array without using traditional spike sorting methods.

### 3.2.2. Estimation of neuronal activity from SR

As previously presented in section 3.1.2, the first PC of the SR array describes the mean neuronal activity (multi-unit PSTH; Fig. 6B, solid line). The second PC describes the difference between the unit with the highest activity and activity of other units (PSTH of unit A minus PSTH of other units; Fig. 6B, dashed line). From the distribution of eigenvalues of the SR array we estimated that most of the signal came from two units only (the first two eigenvalues accounted for 71% of variability and they had considerably higher values from the rest).

### 3.2.3. Spectral spike sorting

The PCA of the transposed SR array enabled identification of spikes features of individual neurons (analogous to spike sorting). The first PC (*pc1*) of the transposed SR array describes the mean distribution of spike features (Fig. 6C, solid line). The single maxima of *pc1* for both parts of SR array indicate that detected units have similar spike waveforms. The second PC (*pc2*) describes dominant temporal changes in distribution of spike features. Thus, *pc2* detects neurons with different temporal firing patterns. The positive and negative parts of *pc2* correspond to spike features of different neurons (Fig. 6C, dashed line). The positions of extremes of *pc2* estimate the centers of spike features distributions for single unit (Fig 7A). The extremes of *pc2* can be mapped to this representation in order to approximate centers of spike clusters (white rectangles in fig. 7A). The same sign of extremes corresponds to the same unit (or more precisely, the same sign of extremes corresponds to the spikes features which are correlated in time). The PCs in panels B and C in figure 7 are the same as the PCs from figure 6C. The PSTHs of both units for 2 kHz sweep is presented in figure 8A.

### 3.2.4. Comparison of pc2 with the difference between units' activity

As stated above, the second PC of the SR array describes the difference between the unit with the highest activity and the activity of other units (PSTH of unit A minus PSTH of other units). Therefore, for the signal from two units only, *pc2* approximates PSTH of unit A minus the PSTH of unit B. As stated above, we found that 71% of variability in this experiment came from two units only. Thus, we can compare *pc2* with PSTH of unit A - PSTH of unit B (Fig. 8B). We find that *pc2* has a very similar profile to the difference of units' activity. Small discrepancies in the relative amplitude and position of extremes indicate activity of additional units not included in this analysis.

## 4. Discussion

In the present study, we constructed spectral representations of extracellularly recorded neuronal data by a discrete mapping of values of spike-waveform features to a time-feature space. This SR encoded information 'when' and 'which' neurons are active without spike sorting while still providing accurate predictive information about a stimuli both in simulated and in real data. Additionally, we were able to estimate single unit activity and assign spike features to individual neurons based on the principal components of the SR array, a technique we call spectral spike sorting.

Standard methods, which rely on spike sorting, have several concerns. First, traditional spike sorting procedures are difficult to automate because an initial human supervised definition of spike clusters or algorithm parameters is required. Furthermore, Wood et al. [25] reported broad disagreement and high error rates in results of spike sorting performed by experts. In contrast, in the SR method no spike clustering is necessary for detecting changes in single unit activity, and spike mapping to the SR is precisely defined (see sample mapping in Appendix).

The improved performance of SR analysis on real data over traditional methods is the result of utilizing additional information from units that are not resolved with spike sorting. Units with spike waveforms with low signal to noise ratios are usually excluded from standard spike sorting analyses. While excluding those units ensures confidence in the results of spike sorting, it reduces the amount of accessible information (e.g. Wessberg et al. [24] showed that prediction accuracy decreases with the number of neurons excluded from analysis). With the SR method, all detected neurons are included. While this increases the available information, it also contributes noise. However, in SR, noise can be easily reduced; indeed, assuming that noise and false spikes are not correlated with the stimulus, the feature selecting algorithms will denoise the signal (e.g. [5]).

In the Results section, we used examples where units had very similar waveforms and low signal to noise ratios. Using 'standards' for spike sorting the units in above examples should be analyzed as multi-unit activity. However, discarding information about neuron identity often results in decreased correct stimulus discriminations, especially when neurons respond to different aspects of a stimulus (such as in the example with simulated data [19]). With SR all available information is utilized, and that increases capacity of neuronal signal decoding. In cases where spike-sorted data is preferred for further analysis, the SR can be applied for analyzing remaining multi-unit activity signal or broadband activity. Furthermore, the SR can be readily expanded to describe features of the local field potential (LFC) and lends itself to integrated analysis of LFPs and spiking activity.

The SR approach also can be readily used for predicting continuous variables (e.g. monkey hand trajectory). In such case, neuronal signal preceding movement (in an appropriate time window) can be mapped to the SR array, and a feature-selecting algorithm can be applied to find changes in SR that are correlated with hand position.

Note that the entire neuronal data analysis with SR can be fully automated. The parameters that were set are easily found automatically (number of PCs to use and bin of PCs). Moreover, we have found that the choice of the threshold for spike detection is not essential for this method. Overall, the SR is relatively unaffected by false spikes or background activity because if the noise is not correlated with the stimulus, then the feature selecting process reduces the noise effectively. Nevertheless, for spike detection, we found more reliable thresholding on rectified neuronal signal convolved with Gaussian kernel (not published data). This operation integrates amplitudes of maximum and minimum of spike waveform, resulting in better separation from noise.

We stress the importance of fully automated analysis of neuronal data in light of the increasing popularity of ensemble recordings. These techniques often use hundreds of electrodes and require rapid on-line analysis and predictions (e.g. [17]). In such approaches, human supervised spike sorting can be a limiting factor.

Interestingly, PCA is not the only valid method for creating and analyzing SR array. Instead of PCA, other dimensional reduction methods (i.e. independent components, factor analysis or wavelets) could be used to describe spike features. Furthermore, other methods for extracting stimulus-related features from SR array could be applied. In the presented example, we used PLS and LDA because of the speed and simplicity of those methods, but SR also could be analyzed with wavelets methods [2], and other classifiers, such as LVQ [12], random forest [1] or support vector machines (e.g. [4]), to name just a few methods tested by us.

The analysis of data in SR is more memory demanding as compared to standard analysis with spike sorting preprocessing. Therefore, for some applications, it could be necessary to reduce the dimensionality of SR (e.g., wavelet compression [3]). Also optimizing/differentiating bin size of time and PC-coordinates can reduce substantially the size of SR array.

**Acknowledgments**

The authors thank Drs Troy A. Hackett and Yoshinao Kajikawa for kindly providing neuronal data, Drs Ronald Coifman, Kenneth D. Harris, Mauro Maggioni and Fred Warner for interesting discussions, Ms Linda Pelechacz and Mr Eyal Kimchi for useful comments and corrections of the manuscript. Special thanks are also due to Dr Mark Laubach for many helpful comments and for supporting the early development of the method.

**References**

[1] Breiman L. Random Forest. University of California Berkeley, Statistics Department Technical Report 2001.
[2] Buckheit J, Donoho DL. Improved linear discrimination using time-frequency dictionaries. Proc SPIE 1995;2569:540-51.
[3] Coifman R, Wickerhauser M. Entropy based algorithms for best basis selection. IEEE Trans on IT 1992;38:713-718.
[4] Cristianini N, Shawe-Taylor J. Support Vector Machines and other kernel-based learning methods, Cambridge University Press, ISBN 0-521-78019-5, 2000.
[5] Donoho DL. De-noising by soft-thresholding, IEEE Trans Inform Theory 1995;41(3):613-627.
[6] Fee MS, Mitra PP, Kleinfeld D. Automatic sorting of multiple unit neuronal signals in the presence of anisotropic and non-Gaussian variability. J Neurosci Methods 1996;69(2):175–188.
[7] Garcia P, Suarez CP, Rodriguez J and Rodriguez M., Unsupervised classification of neural spikes with a hybrid multilayer artificial neural network. J Neurosci Methods 1998;82(1):59–73.
[8] Hackett TA, Preuss TM, Kaas JH. Architectonic identification of the core region in auditory cortex of macaques, chimpanzees, and humans. J Comp Neurol 2001;441(3):197-222.

[9] Harris KD, Henze DA, Csicsvari J, Hirase H, Buzsaki G. Accuracy of tetrode spike separation as determined by simultaneous intracellular and extracelular measurements. J Neurophysiol. 2000;84(1):401-14.

[10] Hulata E, Segev R, Ben-Jacob E. A method for spike sorting and detection based on wavelet packets and Shannon's mutual information. J Neurosci Methods 2002;117:1–12.

[11] Kim KH, Kim SJ. Neural spike sorting under nearly 0-dB signal-to-noise ratio using nonlinear energy operator and artificial neural-network classifier. IEEE Trans Biomed Eng 2000;47(10):1406–1411.

[12] Kohonen T. The self-organizing map, Proc IEEE 1990;78:1464.

[13] Laubach M. Wavelet-based processing of neuronal spike trains prior to discriminant analysis. J Neurosci Method. 2004;134(2):159-168.

[14] Letelier JC, Weber PP. Spike sorting based on discrete wavelet transform coefficients. J Neurosci Methods 2000;101(2):93-106.

[15] Lewicki MS. A review of methods for spike sorting: the detection and classification of neuronal action potentials. Network: Comp in Neural Systems 1998;9(4):53-78

[16] Luczak A, Hackett T, Kajikawa Y, Laubach M. Multivariate receptive field mapping in the marmoset auditory cortex. J Neurosci Methods. 2004, In Press

[17] Nicolelis MAL, Dimitrov D, Carmena JM, Lehew RCG, Kralik JD, Wise SP. Chronic, multisite, multielectrode recordings in macaque monkeys. PNAS 2003;100(19):11041-11046.

[18] Pouzat C, Mazor O, Laurent G. Using noise signature to optimize spike-sorting and to assess neuronal classification quality. J Neurosci Methods 2002;122(1):43-57.

[19] Reich DS, Mechler F, Victor JD. Independent and redundant information in nearby cortical neurons. Science 2001;294: 2566-2568.

[20] Sahani M, Pezaris JS, Andersen RA. On the separation of signals from neighboring cells in tetrode recordings. In Jordan MI, Kearns MJ, Solla SA, eds., Advanced Neural Information Processing Systems, vol. 10, Cambridge, MA, 1998. MIT Press

[21] Shoham S, Fellows MR, Normann RA. Robust, automatic spike sorting using mixtures of multivariate t-distributions. J Neurosci Methods 2003;127(2):111-122.

[22] Schalk G, Carp JS and Wolpaw JR, Temporal transformation of multiunit activity improves identification of single motor units, J Neurosci Methods 2002;114(1):87-98.

[23] Weiss Y. Segmentation using eigenvectors: a unifying view. Proc IEEE Intern Conference on Computer Vision, 1999;975-982.

[24] Wessberg J, Stambaugh CR, Kralik JD, Beck PD, Laubach M, Chapin JK, Kim J, Biggs SJ, Srinivasan MA, Nicolelis MA. Real-time prediction of hand trajectory by ensembles of cortical neurons in primates. Nature. 2000;408(6810):361-365.

[25] Wood F, Fellows M, Vargas-Irwin C, Black MJ, Donoghue JP. Accuracy of manual spike sorting: results for the utah intracortical array. Society for neuroscience 33rd annual meeting, Nov. 8-12, 2003, New Orleans, LA

[26] Zouridakis G, Tam DC. Identification of reliable spike templates in multi-unit extracellular recordings using fuzzy clustering. Comput Methods Prog Biomed 2000;61(2):91–98.

**Appendix**

*Sample calculation of spikes mapping to SR array.*
If the analyzed time window $T = 500$ ms and bin size for time discretization $bin\_t = 2$ ms then the number of columns in SR array $C = T/bin\_t = 250$. If we decide to use two principal components discretised in $M = 100$ intervals (for each component) then number of rows in SR array $R = 2*100$. Thus, values of *pc1* are mapped between 1-100 and values of *pc2* are mapped between 101-200. Values of time are mapped between 1-250. For mapping we can use simple scaling formula:
for *pc1*:  $r1 = round ( M * ( pc1 – min( pc1 )) / ( max( pc1 )-min( pc1 )))$
for *pc2*:  $r2 = round ( M * ( pc2 – min( pc2 )) / ( max( pc2 )-min( pc2 )) + M )$
for time:   $c = round ( t / bin\_t )$
where *r1* and *r2* are the row indexes and *c* is a column index in SR array, *t* is time of spike. Function *round* approximates argument to the nearest integer. Functions *max* and *min* return the biggest and the smallest value respectively.
For example: If spike occurs at time $t = 21.3$ ms and has $pc1 = 0.321$ and $pc2 = -0.12$ (range of PCs: –0.5+0.5), then
$r1 = round ( 100*(0.321+0.5)/(0.5+0.5) ) = 72$
$r2 = round ( 100*(-0.12+0.5)/(0.5+0.5) + 100 ) = 138$
$c = round ( 21.3/2 ) = 11$

thus, this spike would be represented by two 'ones' in SR array: $SR(72, 11) = 1$ and $SR(138, 11) = 1$.

**Figure 1**

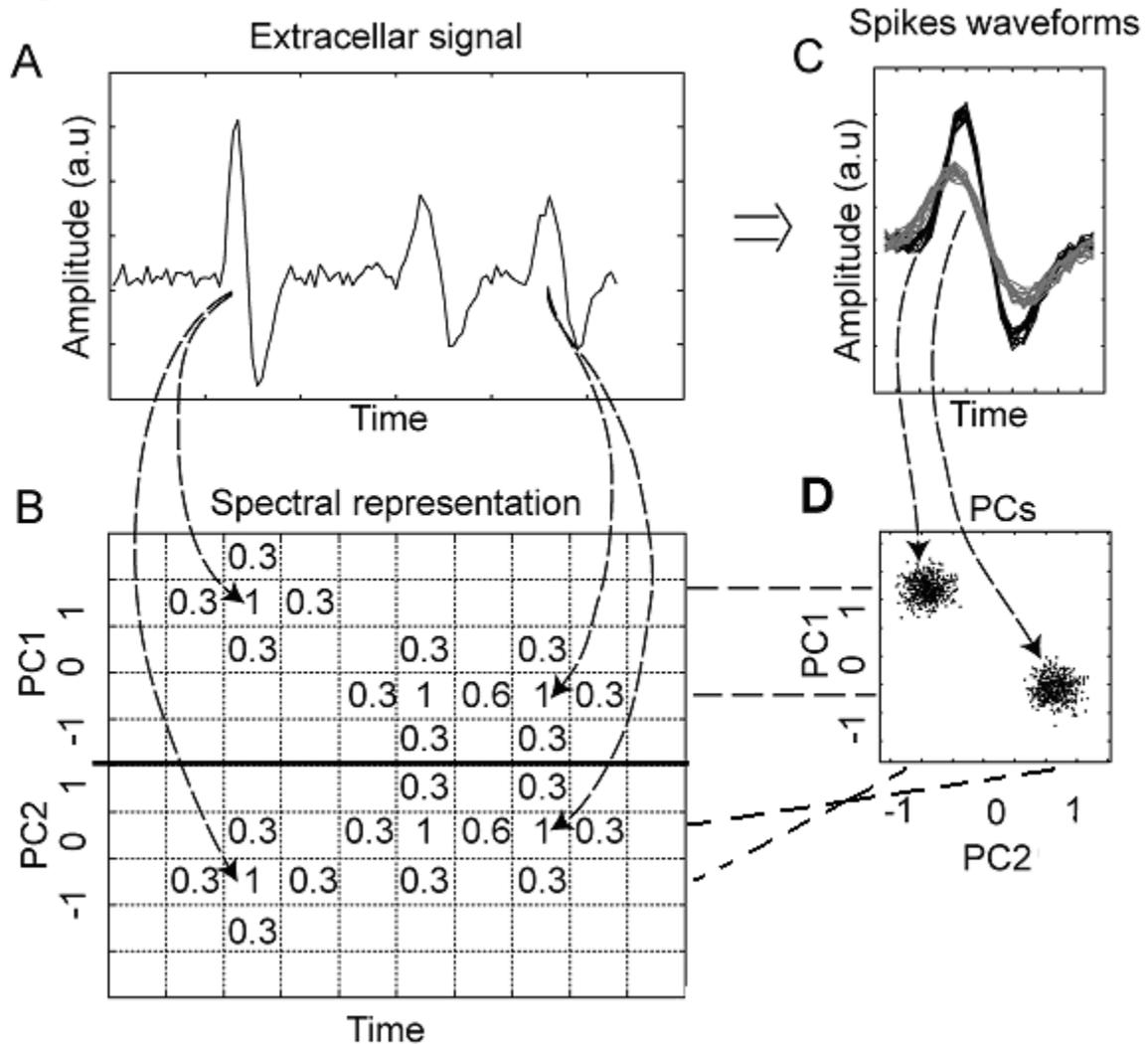

**Figure 1**. Creating spectral representations (SR) of neuronal signals. (A) Extracellular signal with examples of spikes from the two neurons shown in panel C. (B) Spectral representation of neuronal signal from panel A. Every detected spike is represented by a two 'ones' in the SR array where rows correspond to discretised values of two PCs and columns correspond to discretised values of peri-stimulus time (fractions around 'ones' illustrate smoothing of SR array). (C) Example of two distinct spike waveforms. (D) Using PCA the above spike waveforms can be expressed in terms of their first two principal components scores (PC1, PC2).

**Figure 2**

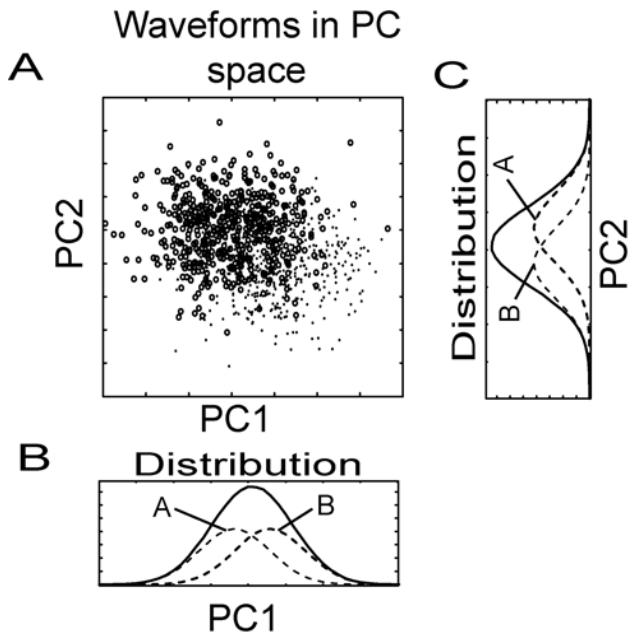

**Figure 2.** (A) Representation of simulated spikes in principal component coordinates (PC1-PC2). Units A and B are denoted by bold and light points respectively. Both units have normal distributions of waveforms features in principal component space. The distributions of unit features are highly overlapping, and therefore difficult to sort. (B) Histogram illustrating distribution of points along PC1 coordinate. Note that maximal number of points is in the middle between centers of the clusters, causing failure of k-means clustering. Dashed lines – simulated distributions of unit A and B; solid line – cumulative distribution (C) Distribution of points along PC2 coordinate.

**Figure 3**

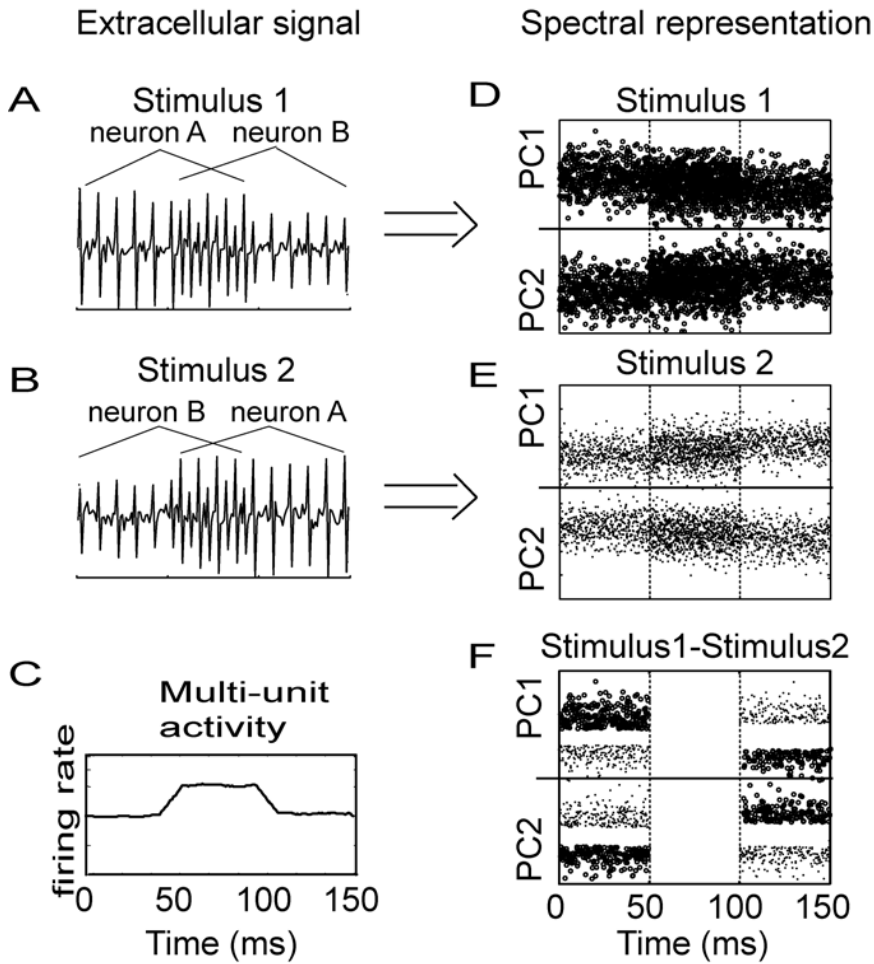

**Figure 3.** Illustration of stimulus discrimination with SR for simulated data. Panels A and B show simulated neuronal responses to the stimuli. When the stimulus 1 is presented (A), the first unit is active between 0-100 ms and the second unit between 50-150 ms. For the stimulus 2 (B) is the opposite response: the first unit is active between 50-150 ms and the second between 0-100 ms. (The spikes amplitude of neuron A is enlarged for visualization of differences between units). (C) For both stimuli the multiunit activity is exactly the same. Both units have very similar spike waveforms (as shown in fig. 2A) making spike sorting practically difficult. (D and E) SRs of neuronal responses to 40 repetitions of the stimuli. (F) Difference between SRs from panel D and panel E indicates that parts of the SR array, that can be used for stimulus discrimination. For illustration purposes, bold points denote neuronal responses to stimulus 1 and light points denote response to stimulus 2. Note, that in this example, where correct spike sorting in principal component coordinates is impossible and multiunit activity does not change with stimulus type, standard neuronal data analysis approaches would fail. While with SR, as shown in panel F, stimulus discrimination can be done automatically and reliably.

**Figure 4**

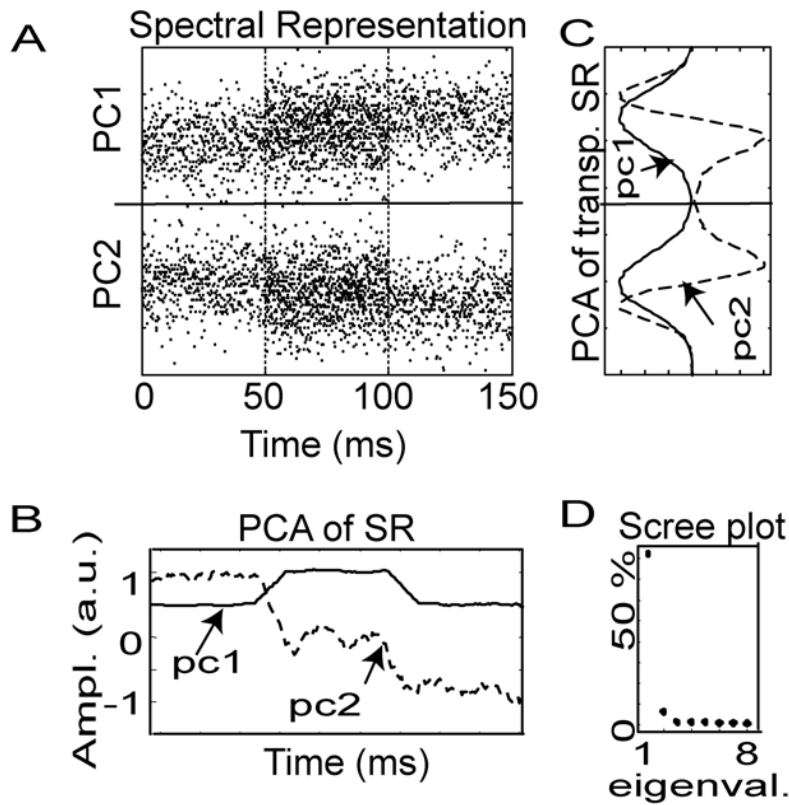

**Figure 4.** Estimating single unit activity from SR and applications of SR for spike sorting. (A) Spectral representation of activity of units A and B. In this representation it becomes evident that we have two units (not visible in fig. 2A). The first unit is active between 0-100 ms, the second between 50-150 ms. This information can be retrieved automatically by decomposing SR with e.g. PCA. (B) The first PC of SR array (*pc1*- solid line) is equivalent to multiunit activity. The second PC of SR array (*pc2* – dashed line) represents difference between units activity. Therefore, when both units are equally active the second PC has values around zero (middle part). The difference between the second PC of SR and the first PC of SR approximates single-unit PSTH (*pc1* + *pc2* $\equiv$ PSTH of unit 1, and *pc1* – *pc2* $\equiv$ PSTH of unit 2). (C) PCA of the transposed SR array. The first PC (*pc1*) represents distributions of values in SR (equivalent to distributions from fig. 2 B and C). The second PC represents difference in distributions between units. The points at which *pc2* is changing sign correspond to borders between clusters of spike features, and extremes of *pc2* correspond to clusters centers. Those points can be mapped back to panel A in fig. 2 for spike sorting purpose (spectral spike sorting). Using SR we can take advantage of difference in tuning properties of neurons to improve spike sorting. (D) Scree-plot illustrating percent of variability captured by PCs in SR array. From the number of PCs necessary to describe variability in SR, number of recorded units can be estimated. In this case only the first two eigenvalues have values significantly above zero, thus number of simulated units is estimated correctly.

**Figure 5**

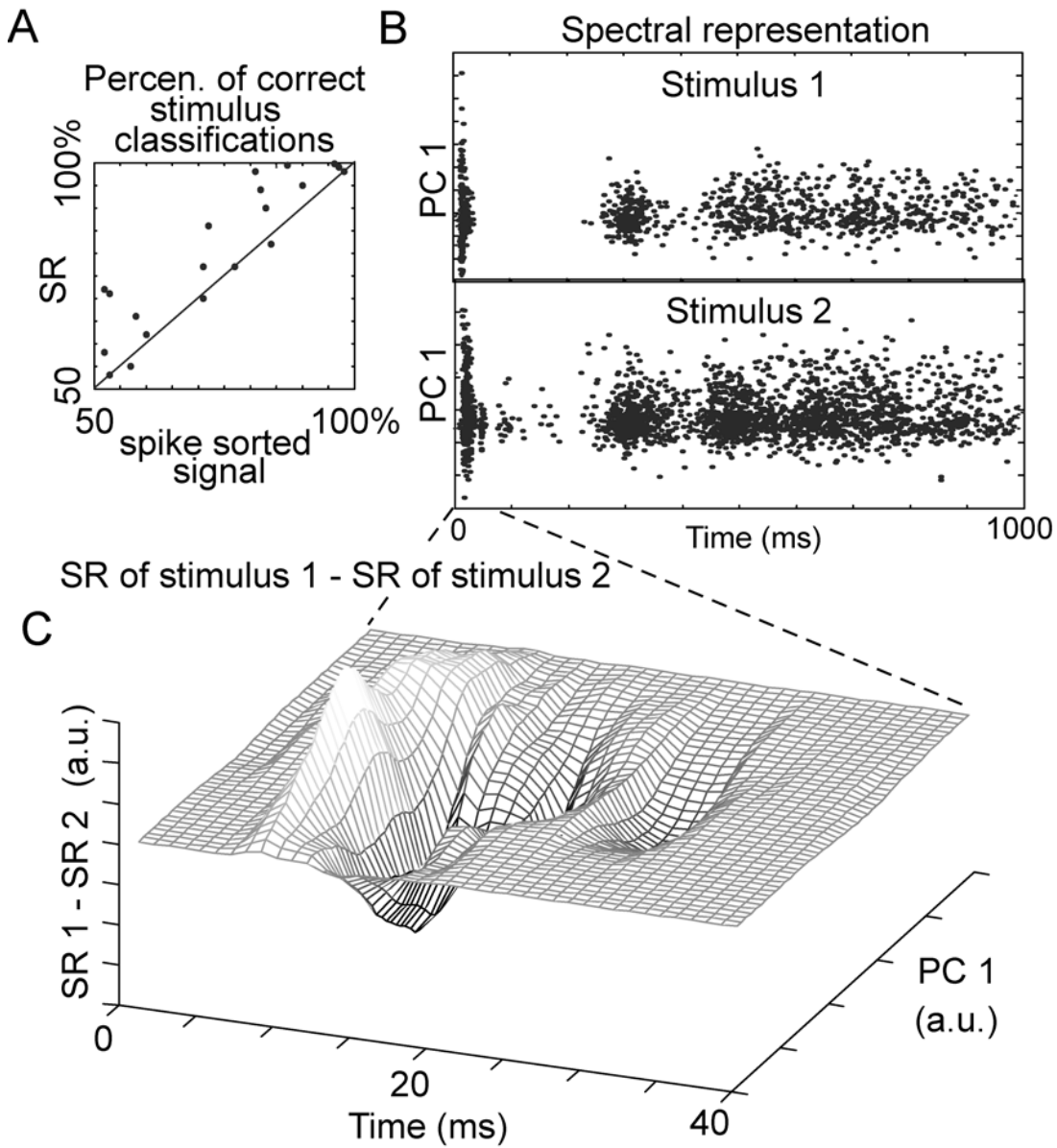

**Figure 5.** Stimulus discrimination with SR using data recorded from marmoset monkey. (A) Percentage of correct stimulus classifications from SR approach vs. using spike-sorted data. (B) The SR of representative neuronal responses to two frequency sweeps (2-10 kHz - upper panel; 6-20 kHz - lower panel). (C) The difference between SRs of neuronal responses to two frequency sweeps (only part of SR array corresponding to 0-40 ms after stimulus onset is shown). The positive peak followed by negative part corresponds to longer response time to 6-20 kHz sweeps. The largest modulations in the lower part of the SR tell us that the neuron with spikes described by low values of PC1, is the most discriminative for presented sweeps. This illustrates extraction of information about tuning properties of a single neuron from SR array, without using spike sorting.

**Figure 6**

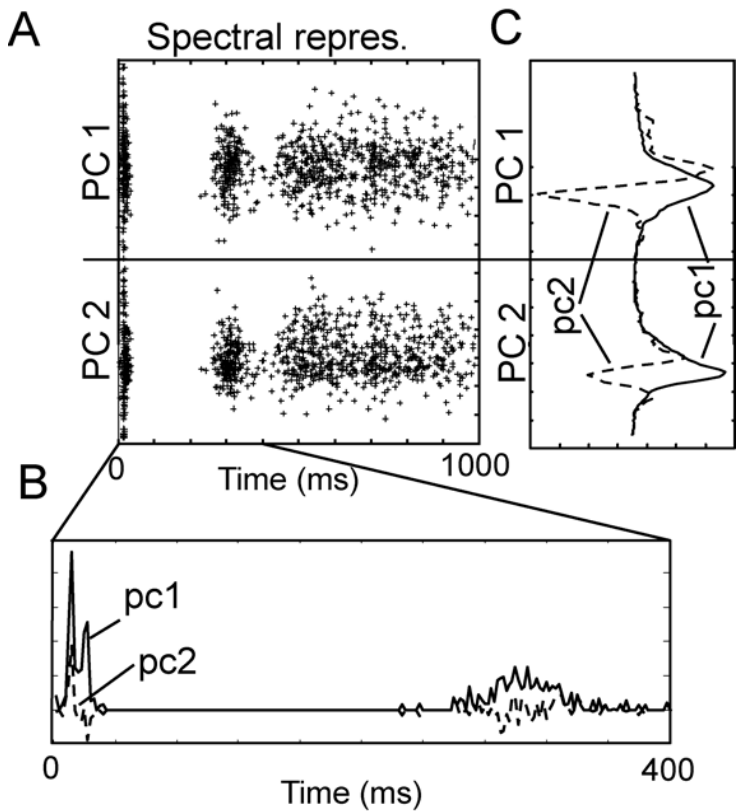

**Figure 6.** (A) The SR of neuronal responses to 2 kHz frequency sweeps. (B) The first PC of the SR array describes the mean neurons activity (solid line). The second PC (dashed line) describes the difference between the unit with the highest activity and the activity of other units. (C) The PCA of the transposed SR array enables identification of spikefeatures of detected units. The first PC of transposed SR array describes the mean distribution of spikes features (solid line). The second PC describes dominant temporal changes in the distribution of spikes features (dashed line).

**Figure 7**

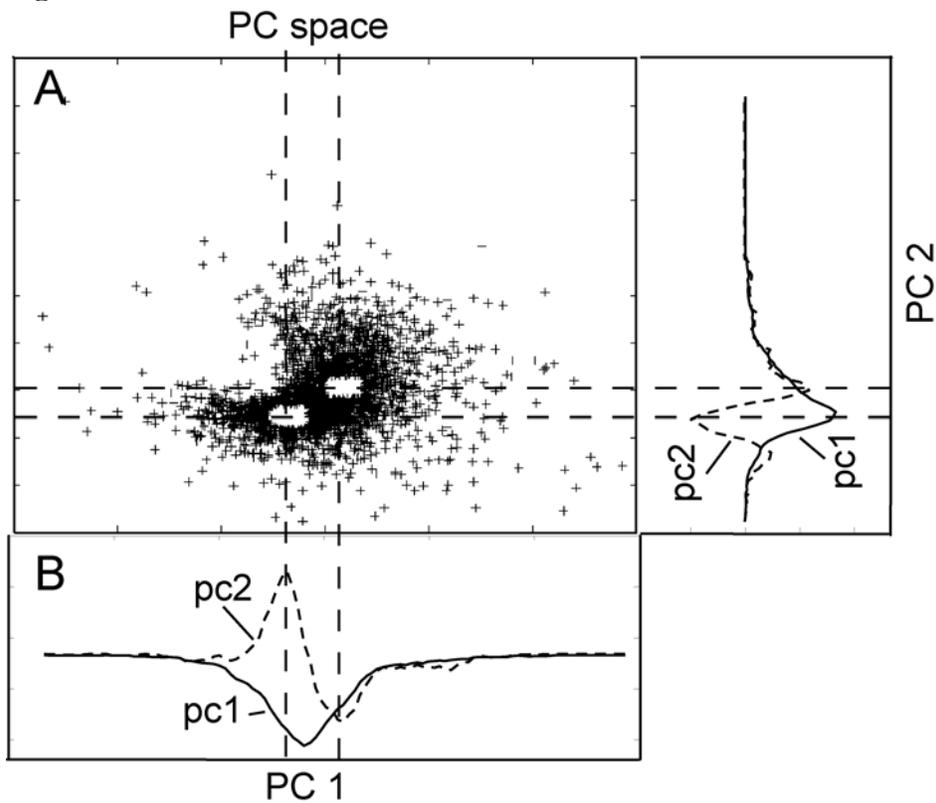

**Figure 7.** (A) Recorded spikes in PC1-PC2 coordinates (white rectangles indicate centers of distributions of two resolved units). (B and C) The PCs of transposed SR array (adaptation of fig. 6 C). The position of extremes of pc2 estimates the centers of spike features distributions for single units. The extremes of pc2 can be mapped to PC1-PC2 coordinates to approximate centers of spike clusters (dashed lines). The same sign of extremes correspond to the same unit (the same sign of extremes correspond to the spikes features which are correlated in time).

**Figure 8**

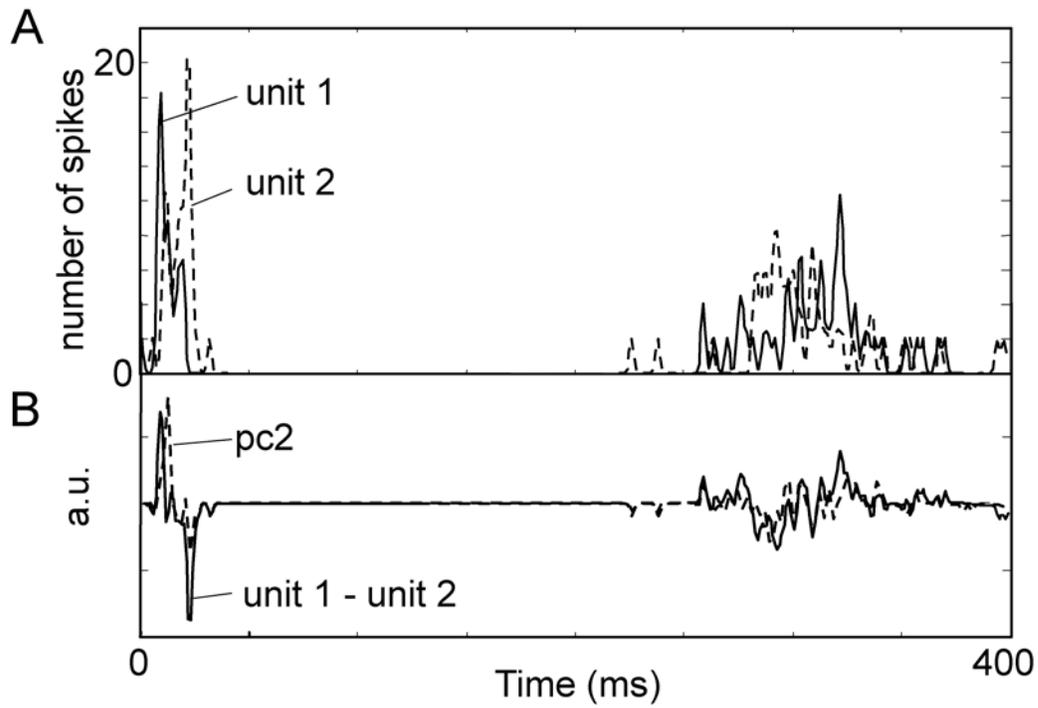

**Figure 8.** (A) The PSTHs of two resolved units. (B) The difference between PSTH of unit 1 and 2 (solid line), and the second PC of the SR array (dashed line, for comparison the pc2 is scaled). The second PC approximates PSTH of unit 1 - PSTH of unit 2. The pc2 has very similar profile comparing to difference of units activity.